%% file: 2016_01_29_corrTreiber.tex
\documentclass[graybox]{svmult}

\usepackage{makeidx}         
\usepackage{mathptmx}       
\usepackage{helvet}         
\usepackage{courier}        
\usepackage{type1cm}        
\usepackage{graphicx}        
\usepackage{epsfig}
\usepackage{amsfonts}
\usepackage{amssymb}
\usepackage{amsmath}
\usepackage{epstopdf}
\usepackage{multirow}
\usepackage[bottom]{footmisc}

\begin{document}

\title*{Calibrating the Local and Platoon Dynamics of Car-following Models on the Reconstructed NGSIM Data}
\titlerunning{Calibrating the Local and Platoon Dynamics of Car-following Models}

\author{Valentina Kurtc and Martin Treiber}
\institute{Valentina Kurtc \at St. Petersburg Politechnic University, Polytechnicheskaya, 29, St.Petersburg\\ \email{kurtsvv@gmail.com}
\and Martin Treiber \at Technische Universtat Dresden, Wurzburger Str. 35, D-01062, Dresden\\ \email{treiber@vwi.tu-dresden.de}}
%
%
\maketitle

\abstract{The NGSIM trajectory data are used to calibrate two car-following models - the IDM and the FVDM. We used the I80 dataset which has already been reconstructed to eliminate outliers, unphysical data, and internal and platoon inconsistencies contained in the original data.We extract from the data leader-follower  pairs and platoons of up to five consecutive vehicles thereby eliminating all trajectories that are too short or contain lane changes.  Four error measures based on speed and gap deviations are considered. Furthermore, we apply three calibration methods: local or direct calibration, global calibration, and platoon calibration. The last approach means that a platoon of several vehicles following a data-driven leader is simulated and compared to the observed dynamics.}

\section{Introduction}
\label{sec:1}
{Nowadays, microscopic traffic data have become more available and
  provides information about thousands of vehicle trajectories. As a
  result, the problem of analysing and comparing microscopic traffic
  flow models with real microscopic data has become more actual. In
  this paper, we consider the NGSIM I80 data set for calibration. Two
  car-following models of similar complexity are studied - the Intelligent-Driver Model (IDM)~\cite{Treiber} and the Full
    Velocity Difference Model (FVDM)~\cite{Jiang}. We apply four
  error measures to investigate the robustness of these models. To
  compare the results with respect to these error measures, the two-sample Kolmogorov-Smirnov test is used. Finally, we compare
    the residual errors of the global and platoon calibration methods to estimate the ratio between inter-driver and intra-driver variations.}

\section{Car-following Models under Investigation}
\label{sec:2}
{Two microscopic car-following models are considered - the IDM  and
  the FVDM. These are formulated as (coupled) ordinary
  differential equations and characterized by an acceleration function
  which depends on the actual speed $v(t)$, the approaching rate $\Delta v(t)=v-v_l$ to the leader, and the gap $s(t)$. Both considered models contain five parameters and are therefore formally equivalent in their complexity.

The IDM is defined by the acceleration function \cite{Treiber}
\begin{equation}
	\dot{v}_{\rm IDM}(v,\Delta v,s)=a\left[1-\left(\frac{v}{v_0}\right)^4-\left(\frac{s^*(v,\Delta v)}{s}\right)^2\right]
\label{eq:03}
\end{equation}
This expression combines the acceleration strategy to reach a desired speed
$v_0$ with a braking strategy that compares the actual gap $s$
  with the dynamically desired gap $s^*(v,\Delta v)=s_0+\max(0,vT+v \, \Delta v/(2\sqrt{ab}))$. A more detailed model description can be found in \cite{Treiber}.

The acceleration function of the FVDM model \cite{Jiang} is as follows
\begin{equation}
	\dot{v}_{\rm FVDM}(v,\Delta v,s)=\frac{v_{\rm opt}(s)-v}{\tau}-\lambda\Delta v
\label{eq:04}
\end{equation}
The model properties are defined by the optimal velocity function $v_{\rm opt}(s)$. 
In this paper we consider it as follows:
\begin{equation}
	v_{\rm opt}(s)=\frac{v_0}{2}\left[\tanh\left(\frac{s}{l_{int}}-\beta\right)-\tanh(-\beta)\right]
\label{eq:05}
\end{equation}
}

\section{\label{sec:3}The Data Set}
{
The Next Generation Simulation (NGSIM) I-80 trajectory dataset
\cite{NGSIM} is considered for calibrating. It was recorded from 4:00
p.m. to 4:15 p.m. on April 13, 2005. The monitored area is
approximately 500 m length and has 6 lanes. The internal and platoon inconsistencies as well as the noise from original data measurements have already been eliminated \cite{NGSIM_report}. Calculating of derived quantities was performed and a smoothing algorithm was proposed in \cite{ThiemannTreiberKesting}. NGSIM data contains information about 3366 vehicle trajectories, that is, for each car we have its current lane position, longitudinal coordinate of its front centre, speed, and acceleration, its length and type (motorcycle, auto or truck), ID of the immediate following and leading vehicle in the current lane.
A great part of the deviations between measured and simulated trajectories can be attributed to the different driving styles, as has been shown in previous works \cite{KestingTreiber}. Microscopic traffic models can easily cope with this kind of heterogeneity because different parameter values can be attributed to each individual driver–vehicle unit. To obtain these distributions of calibrated model parameters, a significant number of trajectories have to be analysed, that is why the NGSIM trajectory data sets are considered in this work.

The consecutive trajectories used for the calibration are extracted by following procedure:
\begin{enumerate}
	\item{Consider trajectory sets of more than 30 s length because the calibration of shorter ones does not sufficiently represent the car-following model properties.}
	\item{Filter out all active and passive lane changes. We do in this way because the car-following models calibrated describe only the longitudinal dynamics.}
	\item{Eliminate the first and last 5 s of the remaining trajectory sets to filter out some inconsistencies. It allows to exclude the influence of not longitudinal effects such as lane changes.}
	\item{Filter out all trajectories on the right most (HOV) and left most (on-ramp) lanes.}
\end{enumerate}
}

\section{Calibration Methodology}
\label{sec:4}
{To find the optimal parameter values of a car-following model with a non-linear acceleration function such as Eq. (\ref{eq:03}) or (\ref{eq:04}), we need to solve a non-linear optimization problem numerically. The MATLAB optimization toolbox is used that provides several algorithms for finding minimum of constrained non-linear multi-variable function. In this case the interior-point algorithm was used.}

\subsection{Simulation Setup and Calibration Methods}
\label{subsec:5}
{We initialize the microscopic model with the empirically given speed and gap, and compute the trajectory of the following car. Then it can be directly compared to the speeds $v^{data}(t)$ and gaps $s^{data}(t)$ provided by the empirical NGSIM data.

Three calibration methods are considered:
\begin{itemize}
	\item{Local or direct calibration: at any time instant, the model's acceleration function is calibrated directly to the observed acceleration. No simulations are needed.}
	\item{Global calibration:  the simulated trajectory of the follower with prescribed leader is compared to the empirical data.}
	\item{Platoon calibration: the dynamics of a platoon of several vehicles following each other with a single data-driven leader are compared to the whole empirical dataset.}
\end{itemize}
}

\subsection{Objective Functions}
\label{subsec:6}
{The calibration procedure aims at minimizing the difference between the measured and simulated dynamic variables. Any quantity which represents aspects of the driving behavior can serve as an objective function, such as the gap $s$, speed $v$, speed difference $\Delta v$, or acceleration $a$. In the following, for global and platoon calibration the errors in the gap $s(t)$ and speed $v(t)$ are used. To assess quantitatively the error between measured and simulated
  data, an objective function is needed. Three types of such measures
are considered. The absolute error measure is given by
\begin{equation}
	S^{abs}=\frac{\sum_{i=1}^n(s_i^{sim}-s_i^{data})^2}{\sum_{i=1}^n(s_i^{data})^2}
\label{eq:06}
\end{equation}
while the relative error measure reads
\begin{equation}
	S^{rel}=\frac{1}{n}\sum_{i=1}^n\left(\frac{s_i^{sim}-s_i^{data}}{s_i^{data}}\right)^2
\label{eq:06a}
\end{equation}
The relative measure is more sensitive to small gaps while the
absolute measure focusses on large gaps. Due to the weighting
  bias of these two methods, we also consider the mixed error measure
  having a more balanced weighting:
\begin{equation}
	S^{mix}=\frac{\sum_{i=1}^n(s_i^{sim}-s_i^{data})^2/|s_i^{data}|}{\sum_{i=1}^n|s_i^{data}|}
\label{eq:07}
\end{equation}
In some papers, the speed instead of the gap is used to measure
  the performance
\cite{PunzoCiuffoMontanino,PunzoMontaninoCiuffo,CiuffoPunzo}. To
compare the calibration results corresponding to different variables,
we also consider the absolute error measure  $S_v^{abs}$ which is
  defined as in Eq.~(\ref{eq:06}) but with the speed as the dynamical quantity.
}

\subsection{Parameter Constraints}
\label{subsec:7}
{The IDM and the FVDM contain five parameters to identify by the
    calibration. To restrict the solution space for optimization to
  reasonable parameter values without excluding possible solutions,
  box constraints are applied. For the IDM, the desired speed $v_0$ is
  restricted to the interval [5, 40] m/s, the minimum distance $s_0$
  to [0, 10] m, the desired time gap $T$ to [-5, 5] s, and the maximum
  acceleration $a$ and the comfortable deceleration b to [0.01, 10]
  $m/s^2$. We explicitly allow negative values for $T$, because some trajectories represent negative time gap values. For the FVDM, the box constraints are [0, 70] m/s for the desired speed $v_0$, [0.05, 20] s for relaxation time $\tau$, [0.1, 100] m for the interaction length $l_{int}$, [0.1, 10] for the form factor $\beta$, and [0, 3] 1/s for the sensitivity parameter $\lambda$.}

\section{Calibration Results}
\label{sec:5}
{Both models have been calibrated for all trajectory pairs or
    platoons satisfying the filtering criteria of
    Sect.~\ref{sec:3}. For the local and global approach, 876
  trajectory pairs were under investigation, whereas for the platoon
  calibration only 251 trajectory sets were studied. For each
  calibration approach, optimal parameter value distributions were
  obtained. Distributions corresponding to different error measures
  were compared by means of the two-sample Kolmogorov-Smirnov test. }

\subsection{Global and Platoon Calibration}
\label{subsec:8}
{Figure \ref{fig:1} visualizes the distributions of the parameter
    values of the IDM (first row) and the FVDM (second row) as obtained from the
    global calibration of all the 876 trajectory pairs with respect to
    the error measure based on the absolute gap differences (Eq.~\ref{eq:06}). Only estimates with residual
    errors below 50 \% are considered. 
\begin{figure}[h]
\begin{center}
\includegraphics[width = 1.0\textwidth]{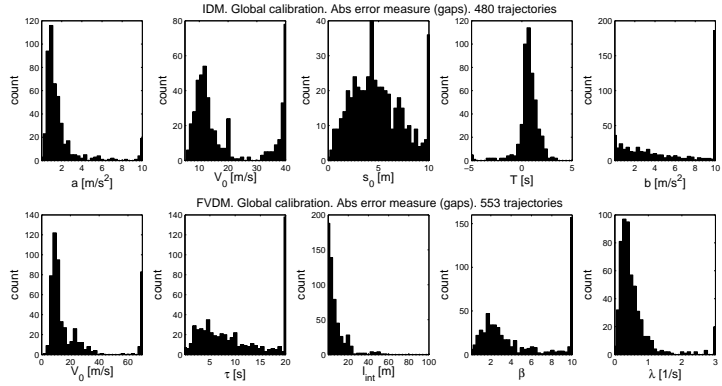}
\caption{IDM and FVDM. Global calibration. Parameter value distributions. Absolute error measure with respect to the gaps.} 
\label{fig:1}       
\end{center}
\end{figure}

To compare distributions obtained with four different measures for each specific model parameter the two-sample Kolmogorov-Smirnov test was used. In this case, the Kolmogorov–Smirnov statistic is
\begin{equation}
	D_{n,n^\prime}=\sup|F_{1,n}(x)-F_{2,n^\prime}(x)|
\label{eq:12}
\end{equation}
where $F_{1,n}$ and $F_{2,n^\prime}$ are the empirical distribution functions of the first and the second sample respectively. The Kolmogorov-Smirnov statistic is in the range from 0.02 (parameter $s_0$, relative and mixed error measures) to 0.27 (parameter $b$, absolute with gaps and absolute with speeds error measures) for the IDM and from 0.02 (parameter $\lambda$, absolute and mixed error measures) to 0.21 (parameter $\lambda$, relative and absolute with speeds error measures) for the FVDM.

For platoon approach we use trajectory sets which contain at least five vehicles following each other. In accordance with filtering rules 251 data sets were considered for calibration. The optimization procedure was evaluated with four error measures as well. Figure \ref{fig:2} presents the parameter values with respect to the error measure based on the absolute gap differences, which estimated errors are less or equal to 100 \% for both models.
\begin{figure}[h]
\begin{center}
\includegraphics[width = 1.0\textwidth]{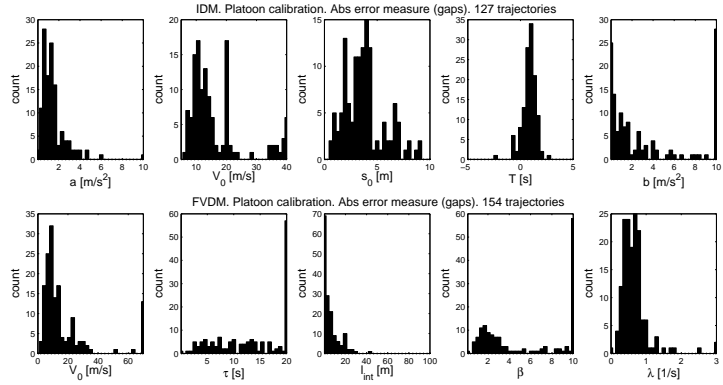}
\caption{IDM and FVDM. Platoon calibration. Parameter value distributions. Absolute error measure with respect to the gaps.} 
\label{fig:2}       
\end{center}
\end{figure}
In case of platoon calibration the Kolmogorov-Smirnov statistic is in the range from 0.05 (parameter $b$, relative and mixed error measures) to 0.38 (parameter $b$, absolute with gaps and absolute with speeds error measures) for the IDM and from 0.05 (parameter $l_{int}$, absolute and mixed error measures) to 0.35 (parameter $l_{int}$ and $\beta [1]$, relative and absolute with speeds error measures) for the FVDM.

Table \ref{table:3} presents the obtained calibration errors. In case of global approach these are from 8.3 \% to 12.5 \%, which is lower than typical error ranges obtained in previous studies \cite{Brockfeld,Ranjitkar,PunzoSimonelli,KestingTreiber}. Platoon method corresponds to higher error values, because it does not allow to distinguish between drivers. These are in the range of 12.8 \% to 32.4 \%.
}

\subsection{Inter-driver and Intra-driver Variations}
\label{subsec:9}
{Let us consider the absolute gap error $\varepsilon_i=s_i^{sim}-s_i^{data}$ for the specific trajectory at time $t_i$. Then we can calculate the variance of this error considering that the mean is equal to zero. It is well-known that variations in driving behaviour come in two forms - inter- and intra-driver variations. In case of global approach the trajectory of one vehicle is calibrated and, thus only intra-driver variation is considered, that is, $\varepsilon^{global}=\varepsilon_{intra}$. The platoon method incorporates several driver styles simultaneously and, as a result takes into account both types of variation $\varepsilon^{s global}=\varepsilon_{intra}+\varepsilon_{inter}$. Assuming no correlation between these two types of errors $cov(\varepsilon_{intra},\varepsilon_{inter})=0$, we can derive the inter-driver variation as follows
\begin{equation}
	Var(\varepsilon_{inter})=Var(\varepsilon^{s global})-Var(\varepsilon^{global})
\label{eq:11}
\end{equation}
Both values in right-hand side of Eq. \ref{eq:11} can be directly calculated. Table \ref{table:4} visualizes the results.

\begin{table}
\begin{minipage}[b]{50mm}
\caption{Calibration errors for IDM and FVDM}
	\begin{tabular}{lcccc}
	\hline
	\multirow{2}{*}{} & \multicolumn{2}{ c }{IDM} & \multicolumn{2}{ c }{FVDM} \\
	\cline{2-5}
	& global & platoon & global & platoon \\
	\hline
	$abs_s$ & 0.098 &  0.256 & 0.097 &  0.239 \\
	$rel_s$  & 0.125 & 0.324 & 0.112 & 0.303 \\
	$mix_s$& 0.111 & 0.296 & 0.105 & 0.279 \\
	$abs_v$ & 0.086 & 0.131 & 0.083 & 0.128 \\
	\hline
	\end{tabular}
\label{table:3}
\end{minipage}
\begin{minipage}[b]{65mm}
\caption{Inter-driver and intra-driver variation. IDM and FVDM}
	\begin{tabular}{lcccc}
	\hline
	\multirow{2}{*}{} & \multicolumn{2}{ c }{IDM} & \multicolumn{2}{ c }{FVDM} \\
	\cline{2-5}
												      & $abs_s$ & $abs_v$ & $abs_s$ & $abs_v$ \\
	\hline
	$Var(\varepsilon^{global}), [m^2]$                                      & 1.74 &  0.35 & 1.83 &  0.32 \\
	$Var(\varepsilon^{super global}), [m^2]$                            & 12.01 & 0.57 & 10.42 & 0.54 \\
	$Var(\varepsilon_{inter}), [m^2]$                                         & 10.27 & 0.22 & 8.59 & 0.22 \\
	\hline
	${Var(\varepsilon_{inter})}/{Var(\varepsilon_{intra})}, [1]$ & 5.9 & 0.6 & 4.7 & 0.7 \\
	\hline
	\end{tabular}
\label{table:4}
\end{minipage}
\end{table}
}

\section{Conclusion}
\label{sec:6}
{The NGSIM trajectory data were used to calibrate two car-following models - the IDM and the FVDM. Four error measures were considered basing on speeds and distances to the leader. Three approaches were used for estimating model parameters - local, global and platoon calibration. During the global calibration the error rates of the models in comparison to the data sets for each model reach from 8.3 \% to 12.5 \%. The global method incorporates only intra-driver variability (a non-constant driving style of human drivers), because it considers only one vehicle following its leader. On the contrary, the platoon approach exploits several drivers simultaneously and, as a result, the inter-driver variation is incorporated as well. Calibration errors in this case are higher and were found to be between 12.8 \% and 32.4 \%. 

The parameter values distributions for the IDM represent negative time gaps $T$ as well. Studying of the empirical trajectories with negative $T$ shows the non-trivial driver behaviour - speed increasing and gap decreasing simultaneously. Such behaviour could be interpreted as failed lane changing.

A significant part of the deviations between
measured and simulated trajectories can be attributed to the
inter-driver variability \cite{KestingTreiber,OssenHoogendoornGorte}
. In this paper we estimated the ratio between inter-driver and
intra-driver variations. It was found between 0.6 \% and 0.7 \% for
calibration according to speeds and from 4.7 \% to 5.9 \%  calibration
with gaps. This ratio is much higher for gaps because, in congested
traffic, the speed is more or less determined by the leading vehicles
while the gap can be chosen freely.

As for benchmarking of car-following models, no model considered in
this study appears to be significantly better. Calibration with four
objective functions and the two-sample Kolmogorov-Smirnov test
demonstrate the same robustness properties of both investigated models.  
}

\input{referenc}

\end{document}

%% file: referenc.tex
%
%
%